
\tolerance 1500
\topskip .75truein

\font\twelverm =cmr12

\def\tt{t^{ons}}
\def\zz{\langle g\rangle}
\def\qs{\lower 5pt\hbox to 23pt{\rightarrowfill}\atop
       {s\rarrow\infty}}
\def\qq{\lower 5pt\hbox to 23pt{\rightarrowfill}\atop
      {q\rarrow\infty}}
\def\q|s|{\lower 5pt\hbox to 23pt{\rightarrowfill}\atop
        {|s|\rarrow\infty}}
\def\ton{\lower 5pt\hbox to 23pt{\rightarrowfill}\atop
       {t\rarrow \tt}}
\def\gon{\lower 5pt\hbox to 23pt{\rightarrowfill}\atop
       {g\rarrow \zz}}

\def\o{\over}
\twelverm
\newdimen\offdimen
\def\offset#1#2{\offdimen #1
   \noindent \hangindent \offdimen
   \hbox to \offdimen{#2\hfil}\ignorespaces}
\parskip 0pt

\def\({\lbrack}
\def\){\rbrack}

\def\Fscr{{\cal F}}

\baselineskip 24pt

\def\aq{\lower 5pt\hbox to 50pt{\rightarrowfill}\atop
       {\hfill a\rarrow\scriptscriptstyle\infty\hfill}}
\def\rin{\lower 5pt\hbox to 50pt{{\rightarrowfill}
     \atop{\hfill r\rarrow\scriptscriptstyle\infty\hfill}}}

\rightline{INFN-ISS 93/9}
\rightline{WIS-93/71/Nov-PH}
\bigskip
\par
\par
\def\tindent#1{\indent\llap{#1}\ignorespaces}
\def\refn{\par\hang\tindent}
\parskip 0pt
\centerline{\bf On the nature of the large-$q$ expansion of
the response for a non-relativistic confined system}
\bigskip
\centerline {E. Pace$^{1}$, G. Salm\`e$^2$}
\centerline {$^1$ Dipartimento di Fisica, Universit\`a di
Roma "Tor Vergata", and INFN, Sezione Tor Vergata,}
\centerline {Via della Ricerca Scientifica, I-00133, Roma, Italia}
\centerline {$^2$INFN, Sezione Sanit\`a, Viale Regina Elena 299,
I-00161 Roma, Italy}
\centerline {A.S. Rinat}
\centerline {Weizmann Institute of Science, Rehovot 76100, Israel}
\bigskip
\par
\bigskip
\par
\vskip 1truein\noindent
\baselineskip 21pt
\par
{\bf Abstract:}

We show the equivalence of a  previously derived exact expression for the
response  of  a  non-relativistic  system with  harmonic  forces  and  an
infinite  sum   of  weighted  $\delta$-functions  corresponding   to  the
spectrum. We forward arguments, indicating that the
Gersch-Rodriguez-Smith $1/q$-expansion of the  response does not converge
and prove that this expansion is an asymptotic series.

 \vskip.8truein\baselineskip
12pt 

\vfil

\eject\
\baselineskip 24pt

\par
In recent years there has been  renewed interest in work
by Gersch, Rodriguez and Smith (GRS). Those authors formally
derived for large momentum transfer $q$,
a $1/q$ expansion of the dominant incoherent part of the total
non-relativistic (NR)
response, $S(q,\omega)$, which is  non-perturbative
in $V\,$$^1$.  This approach has recently been  applied, even if the
underlying
forces are singular and attractive (i.e. confining) $^2$. The incoherent
part of the total reduced response
$\phi(q,y)=(q/m)S(q,\omega)=\phi^{incoh}(q,y)+\phi^{coh}(q,y)$
may thus be expressed as
 $$\phi^{incoh}(q,y)  =  \sum_{k=0}^{\infty}(m/q)^kF_k(y),\eqno(1) $$
where  the energy loss $\omega$   has been
replaced by an alternative variable
$$y=-q/2+m\omega/q\eqno(2)$$
Up  to now  a detailed analysis  of  the  convergence of  GRS
expansion of the response for NR
systems, interacting
either by regular or confining forces,
 is lacking.  In the present note  we consider the possibility that, even
if this  series does  not converge,  a finite number  of terms  may still
describe the response at high $q$ to any desired accuracy, as is the case
for an asymptotic series.

Without  reference to  specific NR
dynamics we consider
the following three equivalent forms for the  response
of a target of mass $m_A$, composed of $A$
constituents with equal mass $m$
$$\eqalignno{ S(q,\omega)&=A^{-1}\sum_n |\Fscr_{0n}(q)|^2
\delta(\omega-q^2/2m_A-E_{n0}) & (3a)\cr &=(\pi A)^{-1}\langle
0|\rho_q^{\dagger}\,\, {\rm {Im}}(\omega+E_0-T-V-i\eta)^{-1}
\,\,\rho_q|0\rangle &(3b)\cr
&=(2\pi A)^{-1}(m/q)\int_{-\infty}^{\infty}ds\,e^{isy}\,e^{isq/2}
\langle 0|\rho^{\dagger}_q\,\, {\rm exp}\left [ {ism\over{q}}\,
(E_0-H)\,\right ] \rho_q|0\rangle, & (3c)\cr}$$
where $\Fscr_{0n}(q)=\langle 0|\rho_q^{\dagger}|n
\rangle$  is the inelastic form  factor between the  ground state
$|0\rangle$ and excited states $|n\rangle$, $E_{n0}$ are intrinsic
excitation energies
without the target recoil energy $q^2/2M_A$, which appears separately
in the energy conserving $\delta$ function in (3a), and $\rho_q $ is the
Fourier transform of the charge density. A formal summation over the
excited states in (3a) produces (3b) in terms of the Green's
function with $T,~ V$ the total kinetic and potential energy. Eq. (3c)
is the Fourier transform of (3b)
and $H = T+V$ the total Hamiltonian of the system. Clearly it is
meaningful only in the  distribution theory.

We now concentrate on systems with confining interactions and shall be
led to questions of equivalence of the different expressions (3). We
shall illustrate our reasoning  on the example of harmonic confining
forces, for which an exact solution for the response  in
terms of a single integration has recently been derived $^3$. It
suffices to treat the case of two particles in a HO well. With
$\beta=(m\Omega/2)^{1/2}$ the inverse oscillator length and
$\alpha=2\beta^2s/q$ one finds for the reduced response
$$\eqalignno{ \phi^{incoh}(q,y)&=\int_{-\infty}^{\infty}
{ds\over{ 2\pi}}\, {\rm exp}(iys)\,
\Phi(\alpha,s)& (4a) \cr
\phi^{coh}(q,y)&=\int_{-\infty}^{\infty}{ds\o {2\pi}}\,
{\rm exp}(iys)\,\,
{\rm exp}
\left\lbrace-{\beta^2s^2\over 4}
\left({{\rm cos}\alpha/2\over{\alpha/2}}\right)^2\right\rbrace
{\rm exp}
\left\lbrace iqs/4
\left({{\rm sin}\alpha\over{\alpha}}+1\right)\right\rbrace
&(4b)\cr}$$
with
$$\eqalignno{\Phi(\alpha,s) &={\rm exp} \left\lbrace -{\beta^2s^2\over 4}
\left({{\rm sin}\alpha/2\over{\alpha/2}}\right)^2\right\rbrace {\rm exp}
\left\lbrace -i{\beta^2s^2\over{2\alpha}}
\left({{\rm sin}\alpha\over{\alpha}}-1\right)\right\rbrace & (5) \cr}$$
The exponents in Eq. (5) are readily seen to be sums of
power series in $\alpha$ and thus permit an expansion of
$\Phi(\alpha,s)$ in series of
$1/q$, except for $q=0$
$$\eqalignno{\Phi(\alpha,s) &=2~\pi\sum_{k=0}^{\infty} (1/q)^k\, h_k(s)
&(6)\cr}$$
When  permitted, interchange  of summation  and term by   term
integration leads to  the GRS sum  (1).  The HO example  is particularly
useful for the discussion of two points:

1) With a priori knowledge
of the equivalence of the expressions (3), can one
directly demonstrate that the integrands in Eqs. (4), analytic
except for $q = 0$, lead to a weighted sum of $\delta$-functions?

2)  The  GRS  expansion  is  formally  generated  through  term  by  term
integration of Eq. (4a) and it consists of a series of regular functions,
(cf.   Ref. 3,  Eq. (3.1)).   The sequence  of the  partial sums  of this
expansion does  not seem  to tend  to a series  of equally  spaced peaks,
required to obtain in the limit Eq. (3a), i.e. a sum of weighted, equally
spaced $\delta$-functions.  In view of this unlikely equivalence, what is
the nature of the GRS series (1)?

We start with the first point and rewrite (4a)
$$\eqalign{
\phi^{incoh}(q,y)&={\rm exp}(-q^2/8\beta^2)
\int_{-\infty}^{\infty}{ds\over{ 2\pi}}\,
{\rm exp}(iys)\,{\rm exp}(iqs/4)\,D(q,s)\cr
D(q,s)&={\rm exp}\left\({q^2\over {8\beta^2}}e^{-2i\beta^2s/q}\right\)
}\eqno(7)$$
The function  $D(q,s)$ is periodic in $2\beta^2s/q$ and
thus permits the Fourier expansion
$$D(q,s)=\sum_{n=-\infty}^{+\infty} d_n(q,\beta)\,
e^{-2in\beta^2s/q}\eqno(8)$$
\noindent
with expansion coefficients
$$d_n(q,\beta)={2\beta^2 \o q}
\int_0^{\pi q/\beta^2}{ds\o {2\pi}}\,e^{2in\beta ^2s/q}\,D(q,s)=
{1\over{2\pi\,i}}\left({{q^2\over{8\beta^2}}}\right)^n\xi_{n+1},
\eqno(9)$$
and
$$\xi_{n+1}\equiv (-1)^{n}\int dz\,
z^{-(n+1)}e^{-z} = {2\pi\,i\over{n!}}\,\,.
\eqno(10)$$
The integration in Eq. (10) is performed on a circle centered in the
origin and of radius $|c|=q^2/8\beta^2$. Eq. (9) can be readily
obtained from Eq. (10) with the substitutions $z=c\,e^{-i\alpha}\,\,
(2\pi\geq\alpha\geq 0)$ and $c = -q^2/8\beta^2$.

In order to show the equivalence of Eqs. (3a) and (3c), we recall that
Eqs. (3c), (4a) and (7) have meaning only
in the framework of  distributions. There we
need the notion of $'$good$'$ functions $g(y)$, i.e. functions which are
everywhere differentiable any number of times, and which together with
their derivatives vanish as $|y| \rightarrow \infty $
faster than any power of $1/|y|~$$^{4a}$.
 For any such $g$,
Eq. (7) implies $$\eqalignno{\int_{-\infty}^{\infty}\phi^{incoh}(q,y)\,g(y)\,dy
&= (2\pi)^{-1}~{\rm exp}(-q^2/8\beta^2)~\int_{-\infty}^{\infty}
e^{isq/4}~D(q,s)\,\widehat {g}(s)\,ds, & (11) \cr}$$ where
$\widehat {g}(s)$ is the   Fourier transform of $g(y)$. We note
moreover
$$\left|\sum_{n=-N}^N d_n(q,\beta)
e^{-2in\beta^2s/q}\right|\,\leq\,\sum_{n=-N}^N
\left({q^2\over{8\beta^2}}\right)^n
{1\over{n!}}\,\leq\,2\,e^{(q^2/8\beta^2)}\eqno(12)$$
and that the right hand side of Eq. (12) is independent of $ N$.
The  Lebesgue dominated convergence theorem $^{4b}$ then permits
inversion of the order of integration and summation of the series in Eq.
(7) and one readily finds
$$\eqalign{ S^{incoh}(q,\omega)&={m\o
q}e^{-(q^2/8\beta^2)}\sum_{n\,=-\infty}^{+\infty}{1\o{n!}}
\left\({q^2\o{8\beta^2}}\right\)^n\delta(\omega-q^2/4m-n\Omega)\cr
&=e^{-(q^2/8\beta^2)}\sum_{n\,=\,0}^{+\infty}{1\o {n!}} \left\({q^2\o
{8\beta^2}}\right\)^n\delta\(y+q/4-(2\beta^2 n/q)\) }\eqno(13)$$
A similar expression can be  derived for the coherent part and one checks
that  the  weights of  the $\delta$-functions  are  the  exact  squared
inelastic form factors  for the HO$^5$.  Notice that  for $\omega>0$,
only  $n\ge 0$, i.e. the actual HO spectrum, contributes to (13).

In spite  of its  utter simplicity, the  above explicit  demonstration of
equivalence is not trivial at all, nor does it serve as a model for other
confining interactions.  Only  for a very limited  number of Hamiltonians
can  one  give  closed  forms   for  the  exact
response, using either  exact Green's function$^6$    or  operator
techniques$^3$. In addition, for no other interaction can the
corresponding function  $D(q,s)$  in (8)  be periodic: it would lead to
an equally spaced spectrum, characteristic for the HO.  In general
a regular  Hamiltonian has a
finite number of unequally spaced discrete  eigenvalues.  In any
case  the proof above  for the equivalence of Eqs.(3a) and (3c) is not
readily seen to be of a general type.

We return  to Eq. (4a).   We already observed  that the integrand  in Eq.
(4a)  permits a  $1/q$-expansion, which  produces the  GRS series  if the
order of summation and integration may be interchanged.  However, we have
not been able to  prove a condition like Eq. (12) for  the partial sum of
the power expansion  of $\Phi(\alpha,s)$, Eq. (6).  A  $finite$ number of
terms in the $1/q$ expansion may always be integrated  term by term, thus
leading to a truncated GRS series and this has been standard practice for
systems with  regular interactions.   A prime  example is  the successful
treatment  of  liquid $^4$He  $^7$.   However,  the crucial  question  is
whether the above interchange is permitted for the $full$ series.  If the
GRS series does not converge, it may  still be an asymptotic one $^8$ and
we now show that this is the case for the HO.

Consider the sequence  $\{S_n(q,y)\}$
$$S_n(q,y)\,=\,\sum_{k=0}^n\,(1/q)^k\,\int_{-\infty}^{+\infty}
h_k(s)\,e^{iys}\,ds.\eqno(14)$$
All integrals  in Eq. (14) are  well defined, since $h_k(s)$  is a linear
combination  of  powers  of  $s$  times  ${\rm  exp}(-\beta^2s^2/4)$  and
therefore $h_k(s)\in~  {\cal L}^1$.   We wish to  show that  the sequence
$\{S_n(q,y)\}$ defines an asymptotic series for $\phi^{incoh}(q,y)$.
To this end we have to show that for any fixed $n$
and  any good function $g(y)$, for any given  $\epsilon > 0$ a
$Q>0$ exists such that, for
$q>Q$ one has $A_n(q)\leq\epsilon$ $^{4a}$ with
$$\eqalign{
A_n(q)~=&~q^n\left|\int_{-\infty}^{+\infty}\phi^{incoh}(q,y)
\,{g}(y)\,dy-\,\int_{-\infty}^{+\infty}S_n(q,y)\,{g}(y)\,dy \right| \cr
{}~=&\,q^n\left|\int_{-\infty}^{+\infty}\sum_{k=0}^{\infty}\,(1/q)^k\,
h_k(s)\,\widehat {g}(s)\,ds\, -\,
\int_{-\infty}^{+\infty}\sum_{k=0}^n\,(1/q)^k\,
h_k(s)\,\widehat {g}(s)\,ds \right|
}\eqno(15)$$
In the second term of the last step of Eq. (15) the order of integrations
has been exchanged  by using the Fubini theorem$^{4b}$.  Eq.  (15) can be
put in the following form
$$\eqalign{
A_n(q)\,=\,{1\over{q}} \left|{\int_{-\infty}^{+\infty}{1\over{(n+1)!}}
{}~\Phi^{(n+1)}(\alpha_o,s)\,\widehat  {g}(s)\,ds\, }\right|},\eqno(16)$$
where $\Phi^{(n+1)}(\alpha,s)\equiv (\partial/\partial x)^{(n+1)}
\Phi(\alpha,s) \,$, $x=1/q =\alpha/(2\beta^2s)$,
and  $\alpha_o=2\beta^2s/q_o$ with $q_o  > q$.  If one can
show that  for any  $\alpha\in (-\infty,+\infty)$
$$\eqalign{\left|{\Phi^{(n+1)}(\alpha,s)}\, \right|\leq {\cal
{P}}_{(n+1)}(s),}\eqno(17)$$
with ${\cal  {P}}_{(n+1)}(s)$ diverging at  most as  a power of  $s$, our
theorem  is  demonstrated.  It  is   easily  seen  that
the derivative  $\Phi^{(n+1)}(\alpha,s)$ is a product of $\Phi(\alpha,s)$
and  combinations of $\ell$-th  order derivatives
( $\ell\leq~n+1$)
of $a(\alpha)={\rm sin}(\alpha/2)/(\alpha/2)$   and   $b(\alpha)=  ({\rm
sin}\alpha/\alpha -1)/\alpha$ with  respect to $\alpha$, with
coefficients
depending  on powers of $s$  with bounded exponents for a given  $n$.

We thus have to study the behaviour of $a(\alpha)$,
$b(\alpha)$ and their derivatives for  $\alpha\in  (-\infty,\,
+\infty)$. Consider separately the two sets
$|\alpha|\leq 1 $  and $|\alpha|> 1$.   In the  range
$[-1,\,1]$ one  may
expand $a(\alpha)$  and $b(\alpha)$  as a
power series  in $\alpha$, which together with the series
defining their  derivatives, converge  uniformly in  the
closed set  $[-1,\,1]$  and their  absolute values   have  there
finite maxima. For $|\alpha|> 1$, the functions  $a(\alpha)$,
$b(\alpha)$ and their derivatives can be expressed  as combinations  of
${\rm  sin}\alpha$,  ${\rm cos}\alpha$  and  powers  of
$1/|\alpha| <1$. One can therefore
find quantities, independent of $\alpha$,  which are
larger than the absolute  values of $a(\alpha)$, $b(\alpha)$  and   their
derivatives for any $\alpha\in  (-\infty,\, +\infty)$. Consequently
(17) holds and our theorem is demonstrated.

We summarize: Although  the GRS expansion for the response  of NR systems
turns out to be  a useful tool, we are not aware of  an actual proof that
the above expansion for systems, interacting through regular or confining
forces is a convergent, or at least  an asymptotic one.  In this note we
have proved  that the latter is  the case for the  $1/q$-expansion of the
response for an harmonic oscillator.

\par

{\bf References}
\bigskip

\refn{$^1$}
H.A. Gersch, L.J. Rodriguez and Phil N. Smith, Phys. Rev. {\bf A5}
(1972) 1547.

\refn{$^2$}
S.A. Gurvitz and A.S. Rinat, Phys. Rev. {\bf C47} (1993) 2901.

\refn{$^3$}
E. Pace, G. Salm\`e and A.S. Rinat, Nucl. Phys., in press.

\refn{$^4$}

a) P. Dennery and A. Krzywicki, $'$ Mathematics for Physicists$'$,
Harper Intl. Edition, Tokio, 1967;

b) A. Kolmogorov and S. Fomin, $'$Elements de la theorie des
functions et de l'analyse fonctionalle$'$, Editions Mir, Moscow, 1977.

\refn{$^5$}
O.W. Greenberg, Phys. Rev. {\bf D47} (1993) 331.

\refn{$^6$}
C. Grosche and F. Steiner, SISSA preprints 1/93/FM; 18/93/FM.

\refn{$^7$}
See, for instance, A.S. Rinat and M.F. Taragin, Phys Rev. {\bf B41} (1991)
4247.

\refn{$^8$}
B. Ioffe, private communication.

\end